\documentclass[runningheads]{llncs}
\usepackage{graphicx}
\usepackage{lipsum} % Just for the demo, you can safely remove it. 
\usepackage[tight,footnotesize]{subfigure}
\usepackage{enumitem} 
\usepackage{float}
\usepackage{multicol,multienum}
\usepackage{breqn}
\usepackage{stfloats}
\usepackage{multirow}
\usepackage{diagbox}
\usepackage{hyperref}
\usepackage{bm}
\usepackage{cite}
\usepackage{graphicx}
\usepackage{tabularx}
\usepackage[ruled,vlined]{algorithm2e} 
\usepackage{booktabs}
\usepackage{marvosym}

\begin{document}
	
	%\title{Reliable Cross-chain-based Federated Learning for Vehicular Edge Networks}
	\title{Scalable and Communication-efficient Decentralized Federated Edge Learning with Multi-blockchain Framework}
	\author{Jiawen Kang\inst{1}, Zehui Xiong\inst{2} (\Letter), Chunxiao Jiang\inst{3}, Yi Liu\inst{4}, Song Guo\inst{5}, \\ Yang Zhang\inst{6}, Dusit Niyato\inst{7},  Cyril Leung\inst{7},  Chunyan Miao\inst{7}}

	\institute{Energy Reserach Institute, Nanyang Technological University (NTU), Singapore 639798, Singapore \and Alibaba-NTU Joint Research Institute, NTU, Singapore 639798, Singapore   \and School of Information Science and Technology,  Tsinghua University,  Beijing 100084, China \and Faculty of Information Technology, Monash University, Clayton VIC 3800, Australia \and Department of Computing,  Hong Kong Polytechnic University (HKPU), Hong Kong and Shenzhen Research Institute, HKPU, Shenzhen 518057, China \and School of Computer Science and Technology, Wuhan University of Technology, Wuhan 430070, China \and 	School of Computer Science and Engineering, NTU, Singapore 639798, Singapore %\\ Corresponding Author: Zehui Xiong,~\email{ zxiong002@e.ntu.edu.sg}
	}

	\maketitle          
	\begin{abstract}
	The emerging Federated Edge Learning (FEL) technique has drawn considerable attention,  which not only ensures good machine learning performance but also solves "data island" problems caused by data privacy concerns. 
	However, large-scale FEL still faces following crucial challenges: (i) there lacks a secure and communication-efficient model training scheme for FEL; (2) there is no scalable and flexible FEL framework for updating local models and global model sharing (trading) management.  To bridge the gaps, we first propose a blockchain-empowered secure FEL system with a hierarchical blockchain framework consisting of a main chain and subchains. This framework can achieve scalable and flexible decentralized FEL by individually manage local model updates or model sharing records for performance isolation. A Proof-of-Verifying consensus scheme is then designed to remove low-quality model updates and manage qualified model updates in a decentralized and secure manner, thereby achieving secure  FEL. 
	To improve communication efficiency of the blockchain-empowered FEL, a gradient compression scheme is designed to generate sparse but important gradients to reduce communication overhead without compromising accuracy, and also further strengthen privacy preservation of training data.  The security analysis and numerical results indicate that the proposed schemes can achieve secure, scalable, and communication-efficient decentralized FEL.
	\end{abstract}

%\author{Jiawen Kang, Zehui Xiong,  Dusit Niyato,~\emph{Fellow, IEEE}, Shengli Xie,~\emph{Fellow, IEEE},\\  Xuemin (Sherman) Shen,~\emph{Fellow, IEEE}, H. Vincent Poor,~\emph{Fellow, IEEE}
%\IEEEcompsocitemizethanks{
%This work was supported in part by WASP/NTU M4082187 (4080), Singapore MOE Tier 1 under Grant 2017-T1-002-007 RG122/17, MOE Tier 2 under Grant MOE2014-T2-2-015 ARC4/15, NRF2015-NRF-ISF001-2277, EMA Energy Resilience under Grant NRF2017EWT-EP003-041, and the National Research Foundation of Korea (NF) Grant funded by the Korean Government under Grant 2017R1A2B2003953. 
%Jiawen Kang, Zehui Xiong and Dusit Niyato are with School of Computer Science and Engineering, Nanyang Technological University, Singapore. (Emails: kavinkang@ntu.edu.sg, zxiong002@e.ntu.edu.sg, dniyato@ntu.edu.sg). Shengli Xie is with School of Automation,  the Guangdong Key Laboratory of IoT Information Technology,  Guangdong University of Technology, China. (Email: shlxie@gdut.edu.cn). %Junshan Zhang is with the School of Electrical, Computer and Energy Engineering, Arizona State University, Tempe, Arizona. (Email: junshan.zhang@asu.edu).
%H. Vincent Poor is with Department of Electrical Engineering, Princeton University, Princeton, New Jersey. (Email: poor@princeton.edu).
%}
%\thanks{}
%}
%\maketitle
%\pagestyle{headings}

%In the battle to fight against COVID-19 pandemic, AI is a very powerful weapon.

\section{Introduction}
With the rapid advancement of Artificial Intelligence, a larger amount of emerging  applications empowered by machine learning technologies significantly enhance the life quality of humans \cite{zhou2019edge}. These applications, such as automatic driving and smart healthcare, utilize advanced machine learning algorithms to train different learning tasks on massive user  data from various edge nodes, e.g., smart phones.  For traditional machine learning approaches,  user data needs to be gathered and centralised in a central server for model training, such as chest CT image analysis for COVID-19  diagnosis.   However, the centralized learning approaches may bring serious data privacy leakage problems. The growing concerns about security and privacy of user data have intensified the demand for new  solutions. A promising machine learning technique named Federated Edge Learning (FEL) is introduced to achieve privacy-preserving model training \cite{konecny2016federated}.  In FEL, the edge nodes collaboratively train a globally shared model by their local data, and only send their local model updates instead of raw data  to a central server \cite{2019kangFL}. The central server gathers all the local model updates to generate an updated global model for the next training iterations. 

Despite that FEL has great advantages for AI-based application with requirements of data privacy protection,  there exist two major challenges for the wide deployment of FEL as follows:  (I)  The central server plays an important role to aggregate  local model updates from edge devices and maintain global model parameters, but is  vulnerable to security challenges, e.g., single point of failure. An unstable central server may result in a system crash. A compromised central server may generate falsified global model to mislead model training and increase system resource consumption.  
(II) There lacks a communication-efficient FEL framework for scalable model training. In the existing FEL framework, edge devices need to frequently upload a large number of local model parameters to the central server for model aggregation, which causes excessive communication overhead and a high demand for network bandwidth \cite{li2020blockchain}.

For the security issues of a single central server,  previous researchers have integrated blockchain into federated learning for secure  model training  \cite{li2020blockchain,kim2019blockchained,8998397}. Kim~\textit{{et al.}} presented a public blockchain-based federated learning framework, in which local model updates are exchanged and verified among miners running energy-hungry Proof-of-Work consensus algorithms  \cite{kim2019blockchained}. Instead of public blockchain,  Lu~\textit{{et al.}} \cite{8998397} proposed a hybrid blockchain framework with an asynchronous learning scheme for secure and efficient federated learning. Similarly,  Li~\textit{{et al.}} \cite{li2020blockchain} designed a decentralized federated learning framework using permissioned blockchain. Although blockchain is an effective way to replace the central server with security guarantee, the process of sharing local model updates among miners brings data privacy leakage challenges to FEL, which is ignored in the existing work. Specifically, recent studies have shown that, even only sharing gradient parameters,  a compromised miner may launch  inference attack that infers features of private training data, even the training data of edge devices, from publicly shared gradients on blockchain \cite{zhu2019deep}. 

For the communication efficiency issues, the existing study presented new consensus mechanisms for blockchain-based FEL to reduce communication cost \cite{li2020blockchain} or developed communication-efficient stochastic gradient descent algorithms \cite{konecny2016federated}, e.g., gradient quantization and encoding \cite{alistarh2017qsgd}.  However, the existing schemes cannot be straightforwardly applied to large-scale FEL because of  high communication-overhead caused by lots of gradients  exchanged between edge devices and a central server (or miners). 
The challenges drive the urgent need of developing secure, decentralized, privacy-preserving and communication-efficient FEL.

To address these challenges,  we first propose a  Blockchain-empowered Federated Edge Learning (BFEL) framework without relying on a trusted centralized server.  In BFEL, a consortium blockchain acting as a trusted and decentralized ledger to  manage model updates from edge devices. To filter out malicious or poisoning model updates, we then propose a  Proof-of-Verifying  (PoV) consensus scheme to collaboratively verify the quality of local model updates among predefined miners. Only the verified model updates can be stored into the block for decentralized federated learning. Since the communication efficiency is significantly important for BFEL, we further integrate a gradient compression scheme  into PoV without lowering learning accuracy. This scheme also relieves inference attack to improve privacy protection of training data.

Moreover, after model training,  learning task publishers can share their models to other entities without enough budget or resources to organize federated learning. For example, a map company can reuse and trade its traffic-prediction training model to vehicles for economic benefit. For the sake of security, the sharing records will be added in the blockchain. However, if both model updates and model sharing records are stored into a single blockchain, this will  result in  larger block size and higher consensus delay. The miners with limited resources cannot synchronize block data in real time.    To avoid this dilemma, we  design further a scalable and flexible framework consisting of a public blockchain as the main blockchain and multiple consortium blockchains as subchains for performance isolation \cite{mutlichain}.
Specifically,  according to data characteristics and service demands (e.g., access control), the model updates from edge devices are respectively  stored on individual subchains named ``Model training subchains".  Meanwhile,   the model sharing records between the task publishers and other entities are stored in a subchain named ``Model trading subchain".

%
%To cope with the above challenges,  we  design a scalable,  secure and decentralized framework consisting of a public blockchain as the main blockchain and multiple consortium blockchains as subchains to increase the flexibility and scalability.
%Specifically, the reputation information is kept as tamper-proof reputation records in a subchain named ``Reputation subchain", and is publicly shared to task publishers for selecting reliable workers for  task allocation. 
%By treating model update records as ``transaction" between the workers and task publishers,  the  model contributions of workers are securely stored on a subchain named ``Training subchain".  %The workers can obtain rewards from the task publishers according to contribution records on the training blockchain.   
%Meanwhile,   the model trading records between task publishers and model users  can be stored in a subchain named ``Model trading subchain" for security consideration.
%The proposed framework simultaneously provides security guarantee to multiple entities on individual subchains  according to the transaction service characteristics and service demands (e.g., access control) \cite{mutlichain}.
%%%This framework can achieve i) data privacy protection by setting access permission on individual subchains and ii) performance isolation through individual consensus algorithms. Each individual subchain maintains its own data locally, and all the subchains are anchored to the main blockchain periodically for publicly verifiable integrity of subchains as well as ensuring interoperability and flexibility.

The main contributions of this paper are summarized as follows.
\begin{itemize}
\item Unlike  single blockchain-based systems,  we design a hierarchical blockchain framework with a main blockchain and multiple subchains to manage model updates and model sharing records in a secure, scalable and flexible manner.

\item For model training subchains, we design a PoV consensus scheme to filter out unreliable model updates by allowing miners to collaboratively verify the quality of  model updates for secure BFEL.

\item  We propose a  gradient compression scheme to improve the communication efficiency of  BFEL without compromising learning accuracy,  and also to enhance privacy preservation by  mitigating inference attacks.
\end{itemize}

\section{Scalable Blockchain Framework for Decentralized FEL} \label{blockchainsystem}

\subsection{Multi-blockchains for Secure Federated Edge Learning} \label{system}

%%%%%%%%%%%%%%%%%%%%%%%%%%%%%%%%%%%%%%%%%%%%%%%%%%%%%%%%%%%%%%%%%%%%%%%%%
\begin{figure}[t]\centering
	\includegraphics[width=8cm,height=7cm]{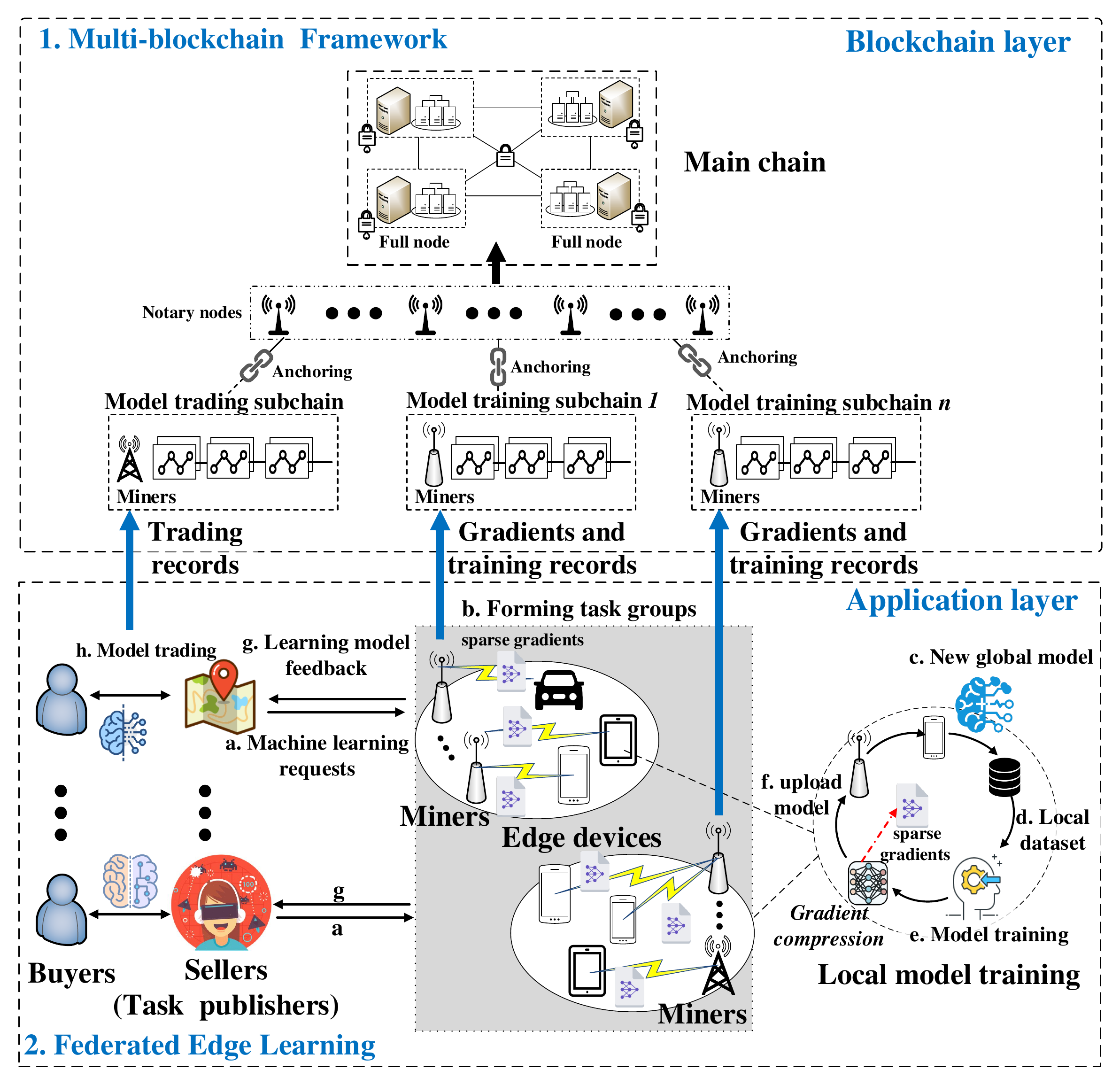}
	\caption{{The proposed federated edge learning framework with multi-blockchain.}}
	\label{systemmodel}
	\vspace*{-6mm}
\end{figure}
%%%%%%%%%%%%%%%%%%%%%%%%%%%%%%%%%%%%%%%%%%%%%%%%%%%%%%%%%%%%%%%%%%%%%%%%%%

%% Òª¸Ä£¬Ì«Ïñ£¡£¡£¡
As shown in Fig.~\ref{systemmodel}, the considered federated edge learning system includes an  application layer and a blockchain layer. In the application layer, each  task publisher, e.g., a map company, sets a learning task (e.g., traffic prediction) and sends the collaborative machine learning request to nearby wireless communication infrastructures, e.g., RoadSide Units (RSUs) in vehicular networks or base stations in cellular networks (Step \textit{a} in Fig.~\ref{systemmodel}) \cite{2019kangFL,liu2020privacy}. These  infrastructures broadcast the learning task to  edge devices with suitable  data (e.g., vehicles or smart phones).  Legitimate  edge devices can join in a task group and act as workers to train the learning task on their local datasets (Step \textit{b}).  Each dataset is generated from personal applications (e.g., navigation services) or collected from surroundings (e.g., sensors on vehicles). Each worker  trains a given global model from its task publisher, and generates local model updates (Steps \textit{c, d, e}). Considering large communication overhead of transmitting local model updates to miners,  a gradient compression scheme is performed to transform the  model updates into compressed model updates with sparse gradients (more details are given in Section~\ref{gradientcompression}). Here, the  miners can be pre-selected RSUs or base stations to establish a consortium blockchain called ``{Model training subchain}". Next, the workers upload their compressed model updates to the miners for model quality evaluation. After executing  Proof-of-Verifying  (PoV) consensus scheme (introduced in Section~\ref{PoV}), the qualified model updates are included into a new block and  stored in a model training subchain (Step \textit{f}). Finally, the workers download the latest block data and calculate a new global model for the next iterations till meeting the accuracy requirements of the task publisher. The final global model is sent back to the task publisher, and the task publisher rewards the workers according to their contributions \cite{2019kangFL}. 
Furthermore, after training, task publishers with high-quality global models  can act as model sellers to trade their models with  model buyers (e.g., drivers) without sufficient cooperating workers or training budget. The model trading records are recorded in a consortium blockchain named ``{Model trading subchain}" for secure storage  (Steps \textit{g, h}).

In the blockchain layer,  blockchains play a significant role in the federated edge learning to provide secure, traceable, tamper-proof data storage (i.e.,  model updates and trading records), which removes the control from a centralized server suffering from security and privacy challenges. However, traditional blockchain systems based on  a single blockchain are not practical and scalable for large-scale FEL because of limited throughput, long consensus delay, and large block size.  Miners in a single blockchain are often overloaded because of constrained resources.  Moreover,  block data  from different services or purposes, e.g.,  model training  and model trading records, should be set different access permission for different entities,  and is stored in isolation  to protect data privacy \cite{mutlichain}. To this end, we propose a multi-blockchain system including  consortium blockchain-based subchains and  a public blockchain-based main chain.

Specifically, by treating model updates as ``transactions" between  workers and task publishers,  local model updates of workers and workers' contributions  are securely stored in their corresponding model training subchains \cite{mutlichain}. Each subchain is only accessible for a task publisher and its participating workers.  Meanwhile, to enable secure and reliable model trading,  the model trading records should be kept as tamper-proof  records in the model trading subchain. Only the task publishers and their model buyers can access and obtain block data in this subchain.  For different subchains,  miners are randomly chosen from communication infrastructures with sufficient computation and storage resources to execute  efficient consensus algorithms (e.g., DPoS and PBFT), respectively. These miners will be changed after each consensus round to reduce the effects of possible collusion among the miners.  The miner selection schemes are out of scope here, but can refer to related work in \cite{zheng2017overview}. 

To efficiently monitor all subchains and miner behaviors, all the subchains should be anchored to the main chain after a time interval for effective governance. To solve the trust problem among blockchains, the block data in the individual subchains can be easily verified by following the notary mechanism in   \cite{mutlichain,wang2019performance}. The main chain periodically stores the Merkle tree root of the block data from different subchains, not the original bock data on the subchains for privacy protection and saving storage resources. This means that the main chain only manages and maintains network addresses of model updates and model trading records. The model buyers can search global models by the latest block data in the main chain, and thus send trading requests to finish the model trading.  
In short, compared with traditional single blockchain-based systems, the proposed framework can achieve: i) data privacy protection by setting access permission on individual subchains and ii) performance isolation through individual consensus algorithms. Each individual subchain maintains its own data locally, and all the subchains are anchored to the main blockchain periodically for publicly verifiable integrity of subchains as well as ensuring scalability and flexibility.

\subsection{Attack Model for Federated Edge Learning}
Although federated learning can solve data privacy issues to a certain extent,  it is subject to new security threats, such as: i) poisoning attack and ii) inference attack. For poisoning attack, malicious edge devices may intentionally send malicious, poisonous or low-quality model updates to poison the global model, thus misleading model training process and  increasing the probability of incorrect learning results \cite{2019kangFL}. 
The poisoning attacks degrade the accuracy of learning tasks, increase the convergence time of the global model, and the probability of erroneous learning results. 
For inference attack, recent studies have shown that a compromised central server (i.e., parameter server) can infer underlying training data by analyzing shared local gradients from edge devices when using gradient-based reconstruction. This intrudes the data privacy of edge devices
illegally and silently \cite{zhu2019deep,Wei2020AFF}. This attack is becoming more serious  because  more entities may obtain shared  gradients in blockchain-based federated learning systems.
Therefore, it is important to defend against the poisoning attack and inference attack for secure and privacy-enhanced federated edge learning \cite{2019kangFL,zhu2019deep}.

%%%%%%%%%%%%%%%%%%%%%%%%%%%%%%%%%%%%%%%%%%%%%%%%%%%%%%%%%%%%%%%%%%
%\begin{figure}[!t]
%	\centering
%	\includegraphics[width=0.7\linewidth]{consensus-d0}
%	\caption{An overview of the  Proof-of-Verifying  Consensus Scheme.}
%	\label{fig-3}
%	\vspace*{-6mm}
%\end{figure}
%%%%%%%%%%%%%%%%%%%%%%%%%%%%%%%%%%%%%%%%%%%%%%%%%%%%%%%%%%%%%%%%%%%%%%%%

\section{ Proof-of-Verifying  Consensus Scheme for Training Subchain}\label{PoV}

In this paper, inspired by the Delegated Proof-of-Stake (DPoS)  consensus algorithm,  we propose an efficient consensus scheme named \textit{ Proof-of-Verifying } (PoV)  that integrates  model updates and quality evaluation into the consensus process, which can defend against poisoning attacks and achieve secure model update and storage. The main steps involved in PoT are as follows.
\begin{itemize}
	\item \textit{Step 1: Initialization}: We adopt an elliptic curve digital signature algorithm and asymmetric cryptography for communication initialization in the system. A Global Trust Authority (GTA)  joins in the proposed PoV to perform identity verification and key manager. Each legitimate entity generates public \& private keys and corresponding certificates for information encryption and decryption after passing  GTA's checking.  
	
	\item \textit{ Step 2: Miner joining:} Communication infrastructures  send joining requests and  submit their resource and identity-related information to the GTA. The GTA will verify the validity of the communication infrastructures based on records of  historical  behaviors.   Only legitimate,  reliable and resource-rich communication infrastructures can be miner candidates to establish subchains. The workers vote for their miner candidates. The candidates with high votes are chosen as delegates and join into  a miner group with a randomly selected leader and other miners acting as verifiers.  The verifiers will execute quality evaluation of  local model updates (described in Step 3). Meanwhile, the leader miner is responsible for aggregating all qualified local model updates and generating pending block.  After each round of consensus, for the sake of safety, the leader and the verifiers will be changed randomly. 
	Similar to DPoS,  all miners should submit a deposit to a shared account under public supervision. If a miner has malicious behaviors during PoV consensus process or causes damage to the global model, the blockchain system will confiscate the deposit and remove the miner. 
		
	\item \textit{Step 3: Quality evaluation of local model updates:} After finishing a local model training process, each worker (i.e., participating edge device)  executes the gradient compression scheme to generate compressed local model updates. More details about the gradient compression scheme are given in Section \ref{gradientcompression}.  
	Then the worker sends its compressed model update to the nearest  miner  on the corresponding model training subchain. This miner (i.e., verifier) first evaluates the quality of the compressed model updates from nearby workers by using a testing dataset. This small testing dataset is verified and provided by the task publisher in each model training subchain, which is considered as a reliable dataset for verifying the training model. Only the qualified model updates, whose accuracy is higher than a given threshold, are picked up to store in a pending block later. The thresholds can be adjusted  according to  security requirements of different task publishers.  In this way, the  model evaluation can prevent poisoning attacks incurred by malicious participants, thus improving security of the proposed  BFEL framework. \cite{li2020blockchain}.
	
	\item \textit{Step 4: Consensus process:}  For mutual monitoring, the verifiers broadcast their model updates and verification results with signatures to each other for double-checking. Each verifier then compares the verification results with those of other miners,  and sends the comparison results as a response to current leader miner for aggregation. The response includes qualified model updates, comparison results, a digital signature, and timestamp.  The leader receives all the qualified local model updates and verifier responses, thus put them into a pending block and broadcasts this pending block to all verifiers. If and only if more than $\frac{2}{3}$  of verifiers agree on the pending block of this round of model updates,  this block data will be added into the model training subchain and synchronized among the all the miners. 
		
	\item \textit{Step 5: Updating training model:} All the workers download the new block data from their corresponding subchains, and calculate the average of all qualified local model updates as their new global models, respectively. The workers will use the new global model for the next training iteration. 
\end{itemize}

%Second, as the compression percentage increases, the number of attack iterations to succeed the CPL attack decreases. This is because a larger portion of the gradients are low significance and are set to 0 by compression. When the attack fails, it indicates that the reconstruction cannot be done even with the infinite(1) attack iterations,

\section{Gradient Compression Scheme for Communication-efficient BFEL } \label{gradientcompression}
In blockchain-empowered federated edge learning,   workers need to send a large amount of gradient information (i.e., local model update parameters)  to miners for aggregating model updates in each training iteration. The workers not only bear large communication overhead, but also suffer  from the inference attack when sharing gradients. However,  previous studies have shown that the sparseness degree of gradient is generally high, so only a few important gradients (i.e., gradients with large absolute values) have a positive effect on the accuracy of the model  \cite{Wei2020AFF}. %However,  previous researches have shown that a large portion of gradients are low significance with little effect on the learning accuracy  \cite{wei2020framework}.  
Inspired by this, we propose a gradient compression scheme to achieve communication-efficient and secure BFEL.
Here, only the important gradients (with large absolute values) are uploaded to the miners to reduce the communication overhead. The importance of a gradient is indicated by its magnitude. Only the  gradients, whose absolute values are larger than a given threshold, are transmitted.  To maintain model performance, the gradient compression scheme utilizes the techniques of momentum correction and local gradient clipping on top of the gradient sparsification to ensure no loss of accuracy \cite{zhu2019deep}. As a result, the gradient compression scheme not only reduces  communication bandwidth problems by  gradient sparsification (i.e., compressing the gradients), but also relieves the inference attack problems by only sharing limited gradient information \cite{zhu2019deep,Wei2020AFF}.
 %DGC also uses momentum factor masking and warm-up training to overcome the staleness problem caused by reduced communication [25]. 

More specifically, the workers only send a part of gradients with large absolute values to their miners.  To avoid information loss caused by gradient sparsification, the rest of gradients are  stored in local buffer space of workers, and   accumulated locally till becoming large enough to be uploaded \cite{zhu2019deep}.  Here,  we use distributed stochastic gradient descent  for iterative updates, and define the loss function to be optimized as follows \cite{zhu2019deep,lin2018deep}:
\begin{equation}
F(\omega ) = \frac{1}{{{D_k}}}\sum\limits_{x \in {D_k}} f (x,\omega ),
\end{equation}
\begin{equation}
{\omega _{t + 1}} = {\omega _t} - \eta \frac{1}{{Nb}}\sum\limits_{k = 1}^N {\sum\limits_{x \in {{\mathcal {B}}_{k,t}}} \nabla  } f\left( {x, {\omega _t}} \right),
\end{equation}
where $F(\omega)$ is the loss function, $f (x,\omega )$ is the loss calculated from data sample $x \in {D_k}$ for workers, and $\omega$ is the weight of the neural network. The learning rate is denoted as $\eta$, and  ${{\mathcal {B}}_{k,t}}$ is a sequence of $N$ mini-batches sampled from ${D_k}$  for the $t$-th round of training ($1 \le k < N$), and $b$ is  size of each local data sample.

Note that the model convergence time  will be affected when the sparsification degree of gradients reaches a large value, e.g., 99\% \cite{zhu2019deep}. To address the convergence problem,  we employ a momentum correction  mechanism proposed in \cite{zhu2019deep,lin2018deep}  to mitigate this effect. Using the momentum correction mechanism,  the accumulated small gradients for each worker converge toward the direction of the gradients with a larger absolute value, thus accelerating the model convergence speed. Moreover, we also apply gradient clipping mechanism to overcome gradient explosion. Specifically, by following \cite{lin2018deep}, the gradient clipping is executed locally before adding current gradients to the previous local gradient accumulation, thus the  gradient explosion problem is alleviated \cite{ref-34,lin2018deep}. 

We prove the gradient compression scheme has no impact on the model convergence as follows \cite{lin2018deep}.   
We define ${g^{(i)}}$ as the $i$-th gradient, and ${u^{(i)}}$ is the sum of the gradients using the optimization algorithm in  \cite{konecny2016federated}. ${v^{(i)}}$ represents the sum of the gradients accumulated in local buffer space, and $m$ is the ratio of the remaining gradients to all gradients. If the $i$-th gradient does not exceed threshold until the $(t-1)$-th iteration and triggers the model update, we have:
\begin{equation}\label{eq-3}
u_{t-1}^{(i)}=m^{t-2} g_{1}^{(i)}+\cdots+m g_{t-2}^{(i)}+g_{t-1}^{(i)},
\end{equation}
\begin{equation}\label{eq-4}
v_{t-1}^{(i)}=\left(1+\cdots+m^{t-2}\right) g_{1}^{(i)}+\cdots+(1+m) g_{t-2}^{(i)}+g_{t-1}^{(i)},
\end{equation}
thus we can update $\omega_{t}^{(i)}=w_{1}^{(i)}-\eta \times v_{t-1}^{(i)}$ and set $v_{t-1}^{(i)}=0$. If the $i$-th gradient is larger than the threshold at the $t$-th iteration, model update is triggered, then we have:
\begin{equation}\label{eq-5}
u_{t}^{(i)}=m^{t-1} g_{1}^{(i)}+\cdots+m g_{t-1}^{(i)}+g_{t}^{(i)},
\end{equation}
\begin{equation}\label{eq-6}
v_t^{(i)} = {m^{t - 1}}g_1^{(i)} +  \cdots  + mg_{t - 1}^{(i)} + g_t^{(i)}.
\end{equation}
Next, we can obtain, 
\begin{equation}
\begin{array}{l}
\omega _{t + 1}^{(i)} = \omega _t^{(i)} - \eta  \times v_t^{(i)}
= {\omega _1}^{(i)} - \eta  \times \big[ {\left( {1 +  \cdots  + {m^{t - 1}}} \right)g_1^{(i)} +  \cdots  + (1 + m)g_{t - 1}^{(i)} + g_t^{(i)}} \big]\\
~~~~~~= w_1^{(i)} - \eta  \times v_{t - 1}^{(i)}.
\end{array}
\end{equation}
Therefore,  the result of using the local gradient accumulation is consistent with the usage effect of the optimization algorithm in  \cite{konecny2016federated}. 
The detailed implementation of the gradient compression scheme is given in Algorithm \ref{al-2} with the following phases:
\begin{itemize}
	\item \textbf{\textit{Phase 1: Local Model Training:}} The workers train their local models on their own  local datasets with  momentum correction and local gradient clipping mechanisms. These mechanisms can solve the learning convergence and gradient explosion problems, respectively. 
\item \textbf{\textit{Phase 2: Gradient Compression:}} Each worker executes the gradient compression process in  Algorithm \ref{al-2} to compress the gradients and upload sparse gradients (i.e., \textbf{only the gradients whose absolute values larger than a threshold are transmitted}) to the nearby miner. Note that the workers send the remaining local gradients in their buffer space to the nearby miner when the local gradient accumulation is greater than the threshold.
\item \textbf{\textit{Phase 3: Gradient Aggregation:}} The miner verifies and  aggregates sparse gradients from local workers. Finally, the qualified gradients from all workers are put into a block, and then both the miners and  the workers can obtain a new global model from the new block data in their corresponding model training subchains.
\end{itemize}
\begin{algorithm}[t]\label{al-2}
	\caption{Gradient compression scheme.}%????
	\LinesNumbered %??????
	\KwIn{A set of workers $\mathcal{N} = \{ {n_1},{n_2},\cdots,{n_i}\} $, $B$ is the local mini-batch size, $D_k$ is the local dataset, $\eta$ is the learning rate, and the optimization function $\mathrm{SGD}$.}%????
	\KwOut{$\omega$.}%??
	Initialize $ \omega_{t}$\;
	${g^k} \leftarrow 0$\;
	\For{$t = 0,1, \cdots $}
	{
		$g_t^k \leftarrow g_{t - 1}^k$\;
		\For{$i = 1,2, \cdots $}
		{
			Sample data $x$ from $D_k$\;
			$g_t^k \leftarrow g_t^k + \frac{1}{{NB}}\nabla f(x;{\omega _t})$\;
		}
	}
	\If{Gradient Clipping}
	{
		$g_t^k \leftarrow \mathrm{Local\_Gradient\_Clipping}\,(g_t^k)$\;
	}
	\ForEach{$g_t^{{k_j}} \in \{ g_t^k\} $ and $j = 1,2, \cdots $}
	{
		$\mathrm{Thr} \leftarrow |\mathrm{Top}\,\rho \%\, of\,\{ g_t^k\} |$\;
		\If{$|g_t^{{k_j}}| > \mathrm{Thr}$}
		{
			Send this gradient to the nearby miner\;
			Send the remaining gradients to the buffer space of  the worker\;
		}
		\ElseIf{When accumulated local gradient $> \mathrm{Thr}$}
		{
			Send this gradient to the nearby miner\;
		}
		All-reduce $g_t^k:{g_t} \leftarrow \sum\nolimits_{k = 1}^N {(\mathrm{sparse}}\, \tilde g_t^k)$\;
		$\omega_{t+1} \leftarrow \mathrm{SGD}\left(\omega_{t}, g_{t}\right)$.
	}
	\Return $\omega $.
\end{algorithm}
	 \vspace*{-8mm}

\section{Security Analysis and Numerical Results} \label{conclusion}

\subsection {Security Analysis}
\subsubsection{\textit{Blockchain-related Issues:}}
The proposed decentralized federated learning framework with multi-blockchains is secure and reliable due to the following reasons: (I) The proposed BFEL framework  can defend against traditional security attacks by standard cryptographic methods including asymmetric and symmetric key-based encryption, and digital signature schemes.
(II) The hierarchical blockchain framework provides flexible authority control. The consortium blockchain-based subchains are established on  authorized infrastructures with different access permissions according to security requirements and configuration. The model training subchains are isolated based on different federated learning tasks.  Only  authorized edge devices and miners can access their corresponding model training subchains.   The model trading subchain is accessible for model buyers and sellers.  The main chain based on public blockchain is open access for all the entities to check and monitor model training records and model training information. This framework enables performance isolation that each individual subchain maintains its own data locally without privacy concerns. All the subchains are anchored to the main blockchain periodically for publicly verifiable integrity of subchains.    
(III) Similar to the DPoS consensus algorithm, the proposed  Proof-of-Verifying  scheme is secure and reliable as long as the number of malicious miners does not exceed $\frac{1}{3}$ of the total number of miners\cite{li2020blockchain}. The malicious miners will be punished and  their deposit confiscated (mentioned in Step 2 of the PoV consensus scheme), which deters the malicious behaviors of miners. 
(IV) The local model update records and model training records are secure because of tamper-proof, decentralization and traceability properties of blockchain technologies \cite{dai2019blockchain,zheng2018blockchain,8813255}.

\subsubsection{\textit{Federated Learning-related Issues:}}
With the help of PoV consensus scheme, both the local model updates and global model updates are reliable and secure for federated edge learning. The reason is that, for the $i$-th round of local model updates, miners  will mutually verify the quality of the local model updates using a given testing dataset, and remove poisonous local model updates  that may damage the global model. Only the high-quality model updates are added into model training subchains to generate a new and reliable global model for the next iteration.  Therefore, the PoV consensus scheme can defend against poisoning attacks and ensure secure decentralized federated edge learning. Moreover, the gradients from workers contains the distribution of local training data.  For inference attacks, the attackers analyze this distribution information and reconstruct the training data according to shared gradients by reverse engineering \cite{zhu2019deep}. Thereby, we can utilize the gradient compression scheme to generate sparse gradients, and upload these gradients to the miners without compromising learning accuracy. Using this approach, we can prevent the attackers from obtaining the complete distribution of local training data, which can reduce gradient privacy issues during decentralized model learning. As a result, the gradient compression scheme not only improves the communication efficiency of BFEL, but also relieves inference attacks caused by gradient leakage problems. 

\subsection{Numerical Results}
We evaluate the performance of the proposed BFEL framework and schemes by using real-world datasets including  MNIST and CIFAR-10.  The datasets are  uniformly divided  into a training set including 70\% data  and the rest data is included in a test set. We implement the proposed BFEL framework  using Pytorch, PySyft, and a blockchain platform named EOSIO with DPoS scheme \cite{2019kangFL}. The experiment is conducted on a virtual workstation with the Ubuntu 18.04 operating system, Intel (R) Core (TM) i7-4500U CPU, 16GB RAM, 512GB SSD. There exist 2 task publishers, 22 miners, 20 workers, and also a model trading subchain and 2 model training subchains in the simulation. All of the subchains apply the DPoS scheme as their consensus algorithms.

%%%%%%%%%%%%%%%%%%%%%%%%%%%%%%%%%%%%%%%%%%%%%%%%%%%%%%%%%%%%%%
%\begin{figure}[!t]
%	\centering
%	\includegraphics[width=0.5\linewidth]{accuracy}
%	\caption{The accuracy of BFEL framework with different gradient thresholds  $\rho$.}
%	\label{fig-5}
%		 \vspace*{-3mm}
%\end{figure}
%%%%%%%%%%%%%%%%%%%%%%%%%%%%%%%%%%%%%%%%%%%%%%%%%%%%%%%%%

In our Blockchain-based Federated Edge Learning (BFEL) framework, the gradient compression scheme plays an important role for system performance. We first evaluate effects of a hyperparameter  $\rho$  (i.e., the threshold of gradient absolute value in Algorithm 1) for the BFEL. A simple Convolutional Neural Network (CNN) network (i.e., CNN with 2 convolutional layers followed by 1 fully connected layer) is used to perform the classification tasks on MNIST and CIFAR-10 datasets, respectively. The pixels in all datasets are normalized into range of [0,1]. In the simulation, we take a model training subchain with 10 workers and 11 miners as an example. The learning rate is  $\eta = 0.001$, and the training epoch is  $E= 1000$. The mini-batch size is  $B = 128$, and  $\theta$ is set  as 0.05.
We compare the performance of different  $\rho$ thresholds for the learning accuracy, and thus find out the best threshold of the gradient compression scheme in our simulation.

\begin{figure}[htbp]
	\centering
	\begin{minipage}[t]{0.48\textwidth}
		\centering
		\includegraphics[width=6cm]{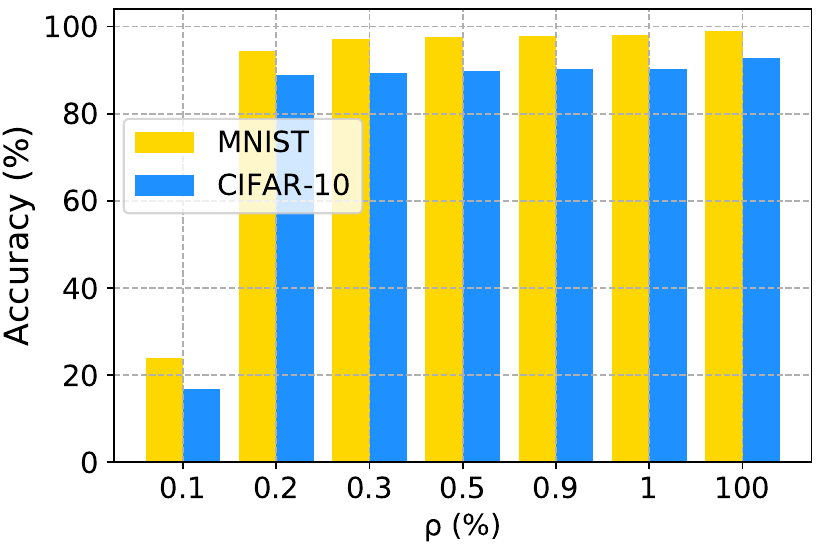}
		\caption{The accuracy of BFEL framework with different gradient thresholds  $\rho$.}
			\label{fig-5}
	\end{minipage}
	\begin{minipage}[t]{0.48\textwidth}
		\centering
		\includegraphics[width=6cm]{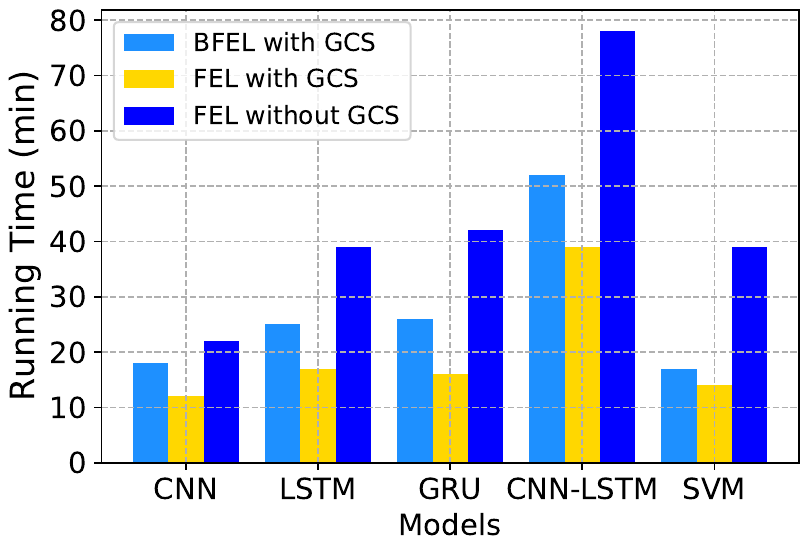}
		\caption{Comparison of communication efficiency in different scenarios and models.}
			\label{fig-8}
	\end{minipage} \vspace*{-3mm}
\end{figure}

Specifically,  $\rho$ takes value from the set  $\{0.1,0.2,0.3,0.5,0.9,1,100\}$ to carry out simulation on the  MNIST and CIFAR-10 datasets to observe the best threshold of the gradient compression scheme. As shown in Fig. \ref{fig-5}, we observe that the larger $\rho$ leads to the better accuracy performance of the proposed framework. For the MNIST task, the results demonstrate that the accuracy is 97.25\%  when $\rho = 0.3$, and the accuracy is 99.08\% when $\rho = 100$. This means that although the gradient size has been raised more than 300 times as compared with $\rho = 0.3$,  the learning accuracy is only improved 1.83\% than that of $\rho = 0.3$. Furthermore, we observe a trade-off between the gradient threshold and accuracy. Therefore, to achieve the trade-off between the gradient threshold and the learning accuracy, we set $\rho = 0.3$ as the best threshold of the gradient compression scheme. 

%%%%%%%%%%%%%%%%%%%%%%%%%%%%%%%%%%%%%%%%%%%%%%%%%
%\begin{figure}[!t]
%	\centering
%	\includegraphics[width=0.5\linewidth]{runningtime}
%	\caption{Comparison of communication efficiency in different scenarios and models.}
%	\label{fig-8}
%		 \vspace*{-3mm}
%\end{figure}
%%%%%%%%%%%%%%%%%%%%%%%%%%%%%%%%%%%%%%%%%%%%%%%%%%

%%%%%%%%%%%%%%%%%%%%%%%%%%%%%%%%%%%%%%%%%%%%%%%%%%%%%%%%%%%
%%%%%%%%%%%%%%%%%%%%%%%%%%%%%%%%%%%%%%%%%%%%%%%%%%%%%%%%%%

%We then compare the communication efficiency of the BFEL framework with the gradient compression mechanism (GCS) with the traditional centralized FEL framework with or without GCS.
%We apply typical CNN, LSTM \cite{ref-39},  Gate Recurrent Unit (GRU) \cite{ref-41},  CNN-LSTM \cite{ref-40},  and Support Machine Vector (SVM) \cite{ref-43} methods with an identical simulation configuration.
For the communication efficiency of  the BFEL framework, we compare the BFEL framework with the Gradient Compression Scheme (GCS) with the traditional centralized FEL framework with or without GCS.
We apply typical CNN, Long Short-Term Memory (LSTM),  Gate Recurrent Unit (GRU),  CNN-LSTM,  and Support Machine Vector (SVM) methods with an identical simulation configuration. For these methods, CNN is running on MNIST dataset to execute an image classification task, and the rest of methods are running on a power demand dataset with time series data to perform power consumption prediction task \cite{powerdataset}. The gradient threshold $\rho$ of the GCS is set as 0.3. Similar to DPoS  in EOSIO platform, the consensus time of PoV  scheme in each  round  is set as 0.5 seconds for the BEFL framework \cite{eosio}. 
Considering the communication overhead of each round as a fixed  value, we compare the running time of the above methods in three scenarios (i.e., BFEL with GCS,  FEL with or without GCS) to indicate the communication efficiency.  As shown in Fig. \ref{fig-8}, we observe that the running time of FEL framework with GCS is less 50\% than that of FEL without GCS. The reason is that GCS can reduce the number of gradients exchanged between the workers and the cloud aggregator. Since there exists delay caused by PoV scheme in BFEL,  the running time of BFEL framework with GCS in different scenarios is  higher than that of FEL with GCS, but much lower than that of FEL without GCS. Moreover, the BFEL framework with GCS can defend against poisoning attacks by the PoV  scheme and remove the centralization security challenges by blockchain technology.  Furthermore, GCS can compress the gradient size by 300 times with almost no reduction in accuracy.   Therefore, the proposed BEFL framework is more secure, communication-efficient and practical in real-world applications.

\section{Conclusions} \label{conclusion}
In this paper, we propose BFEL, a scalable, communication-efficient, blockchain-based  framework for federated edge  learning. First, we introduce a hierarchical blockchain framework with multiple blockchains to manage training models and model trading records in a scalable and flexible way. Second, we propose a  Proof-of-Verifying  consensus scheme to defend against poisoning attacks and ensure reliable federated edge learning. Third, a gradient compression scheme is presented to reduce communication overhead and achieve communication-efficient federated edge learning. We evaluate the performance of the proposed framework and schemes on real-world datasets with different typical machine learning methods. Security analysis and numerical results indicate that the proposed framework not only ensures secure, scalable federated learning, but also achieves communication-efficient federated edge learning. \\

{\textbf{Acknowledgments.}}
 This research is supported by the National Research Foundation (NRF), Singapore, under Singapore Energy Market Authority (EMA), Energy Resilience,  NRF2017EWT-EP003-041,  NRF2015-NRF-ISF001-2277, Singapore NRF National Satellite of Excellence, Design Science and Technology for Secure Critical Infrastructure NSoE DeST-SCI2019-0007, A*STAR-NTU-SUTD Joint Research Grant on AI for the Future of Manufacturing RGANS1906, Wallenberg AI, Autonomous Systems and Software Program and NTU under grant M4082187 (4080), and Joint NTU-WeBank Research Centre on FinTech (Award No: NWJ-2020-004), NTU, Singapore,  Alibaba Group through Alibaba Innovative Research (AIR) Program and Alibaba-NTU Singapore Joint Research Institute (JRI), NTU, Singapore, and also the National Natural Science Foundation of China (Grant 61872310) and the Shenzhen Basic Research Funding Scheme (JCYJ20170818103849343), the Open Fund of Hubei Key Laboratory of Transportation Internet of Things, China (No. WHUTIOT-2019005), and the Open Research Project of the State Key Laboratory of Industrial Control Technology, Zhejiang University, China (No. ICT20044).

%guosong?Please list both ?Department of Computing, The Hong Kong Polytechnic University? and ?Shenzhen Research Institute, The Hong Kong Polytechnic University? as my affiliation, and include the following information in ACK: ?This research was financially supported by the National Natural Science Foundation of China (Grant 61872310) and the Shenzhen Basic Research Funding Scheme (JCYJ20170818103849343).?

%zhangyang?the Open Fund of Hubei Key Laboratory of Transportation Internet of Things, China (No. WHUTIOT-2019005) the Open Research Project of the State Key Laboratory of Industrial Control Technology, Zhejiang University, China (No. ICT20044)

%dusit: This research/project is supported by the National Research Foundation (NRF), Singapore, under Singapore Energy Market Authority (EMA), Energy Resilience,  NRF2017EWT-EP003-041, Singapore NRF2015-NRF-ISF001-2277, Singapore NRF National Satellite of Excellence, Design Science and Technology for Secure Critical Infrastructure NSoE DeST-SCI2019-0007, A*STAR-NTU-SUTD Joint Research Grant on Artificial Intelligence for the Future of Manufacturing RGANS1906, Wallenberg AI, Autonomous Systems and Software Program and Nanyang Technological University (WASP/NTU) under grant M4082187 (4080), and Joint NTU-WeBank Research Centre on FinTech (Award No: NWJ-2020-004), Nanyang Technological University, Singapore.

%\bibliography{myreference}

\end{document}